\documentclass[fleqn,twoside]{article}
\usepackage{espcrc2}
\usepackage{amssymb}

\usepackage{graphicx}
\usepackage[figuresright]{rotating}

\newcommand{\AmS}{{\protect\the\textfont2
  A\kern-.1667em\lower.5ex\hbox{M}\kern-.125emS}}

\title{Status report of the ANTARES project}

\author{T. Montaruli\address[Bari]{Physics Department and I.N.F.N., 
        University of Bari, \\ 
        I-70126 Bari, Via Amendola 173, Italy},
        on behalf of the ANTARES Collaboration}
       
\begin{document}

\begin{abstract}
\vspace{1pc}
The ANTARES project aims at the construction of an underwater neutrino
telescope at the scale of 0.1 km$^2$ 2400 m deep in the Mediterranean Sea. 
After a 4-year R\&D program, the 
ANTARES project has entered the construction phase which will be
concluded by the end of 2004.
The current status of the project is reported.
\end{abstract}

\maketitle

\section{The ANTARES project}

Neutrino telescopes of huge dimensions can investigate
the far Universe and the Very High Energy (VHE) cosmic ray sources.
Neutrinos are unique `probes' to investigate regions at larger distances
than 50 Mpc, because they are weakly interacting 
and point backwards to their source. 
Large neutrino telescopes are designed to detect the Cherenkov
light emitted by charged particles in deep 
water/ice using an array of phototubes (PMTs)
enclosed in high pressure resistant glass spheres called
optical modules (OM).
Upward going muons produced by neutrinos
having crossed through the Earth,
are recognized as products of $\nu$ interactions in the instrumented 
region or close to it. Neutrino telescopes are expected to 
measure the muon direction and energy, while
distinguishing the neutrino signal from the atmospheric muon background,
which is orders of magnitude larger. 
PMT pulse times and amplitudes allow us to calculate 
these quantities.
  
The ANTARES (Astronomy with a Neutrino Telescope and Abyss environmental 
RESearch) project \cite{Proposal}, 
is complementary to already existing experiments, such
as AMANDA \cite{AMANDA} and Lake Baikal \cite{Baikal}, 
in covering the whole sky looking for neutrinos
of astrophysical origin.  
The ANTARES collaboration involves physicists and astrophysicists, sea
science experts and engineers from France, Italy, the Netherlands, Russia,
Spain and the United Kingdom. The project started in 1996.
During a 4 year R\&D phase, extensive campaigns of measurements of 
environmental parameters have lead to the selection of the detector site
located 40 km off the Toulon coast (South France) at a depth 
of 2400 m. 
Thanks to the rotation of the Earth, 
this location (42$^{\circ}$ 50'N, 6$^{\circ}$ 10'E) has an
efficient sky coverage of about 3.5$\pi$ sr, with approximately a 
0.5$\pi$ sr overlap 
with AMANDA, and allows to survey the Galactic Center, not accessible
to AMANDA.

\section{The R\&D phase}

Since 1996, several measurement systems have been designed and operated
in order to locate the site where the deployment of the detector is
possible and to study the sea-water properties which largely determine
the design of the detector.

Surveys have qualified the sea bed for detector deployment;
observed currents, on average $\sim 6$ cm/s with maximum peaks of 19 cm/s,
are taken into account to allow long term continuous 
operation.
When exposed to sea water, optical module surfaces are fouled by 
living organisms and
by sediments. These processes reduce light collection efficiency.  
Results from about 8 months of measurements with a system of PIN diodes
illuminated by a blue LED at angles between 50$^{\circ}$ to 90$^{\circ}$ 
from upward vertical, have shown that at $90^{\circ}$ fouling induces
a light loss of about 2\,\% per year. Consequently,
ANTARES PMTs will be oriented downwards.
Water transparency determines light detection efficiency, while
the amount of scattered photons affects reconstruction 
and hence pointing capabilities.
The photon arrival time distributions for a pulsed isotropic LED source
at wavelengths of 370 nm (UV) and 470 nm (blue)
have been measured at 2 different distances of a 1'' PMT from
the source (24 m and 44 m). These distributions are
used in simulations including the PMT angular response 
to develop and test reconstruction techniques, described in detail in
\cite{Carmona}.
For a distance of 24 m, 95\% of the emitted photons are collected 
by the detector within 10 ns (30 ns) for blue (UV) light.
It is found that for blue light the absorption length
is $\sim 55$ m and the effective scattering length (the
scattering length divided by $1-\langle\cos\theta\rangle$, where
$\theta$ is the scattering angle) is in the 300 m range.
The scattering contribution is negligible in sea water compared to ice.

Trigger logic and electronics, track reconstruction
and background rejection must take into account the presence of
background light due to the $\beta$-decay of $^{40}$K present in sea
water. This light produces an almost constant counting rate
of about 60 kHz on a 10 inch PMT (much higher
than the background rate for AMANDA at the South Pole). 
Moreover, short bioluminescence
bursts of MHz counting rate with few ms rise-time over a few seconds 
have been measured.
These induce on average less than 5\%
dead time on PMTs with negligible correlation between storeys.

From Nov. 1999 to June 2000, a demonstrator string was deployed at a
depth of 1100 m and connected to shore with a 37-km electro-optical cable.
The string had a structure different from the current design, 
containing couples of glass spheres separated by 15 m.
Near the bottom 7 OMs were equipped with PMTs oriented 
horizontally.
More than 50000 of 7-fold coincidences were recorded and zenith angles
were reconstructed using recorded times and amplitudes.
We reproduced the shape of the zenith angular
distribution of atmospheric muons reasonably well, in spite of the single 
string and the small number of PMTs. After reconstruction,
the simulation which is compared to the data includes 
a $\sim 50\%$ contribution of multiple muons (due to the shallow depth
of the string).

The demonstrator string allowed us to test the ANTARES relative and absolute 
positioning with
a system of acoustic rangemeters, compasses 
and tiltmeters.
Relative distances between 2 elements were measured with  
5 cm accuracy while absolute positioning was obtained with $\sim 1$ m
accuracy.
\begin{figure}[htb]
\vspace{-11pt}
\includegraphics[width=18pc, height=13pc]{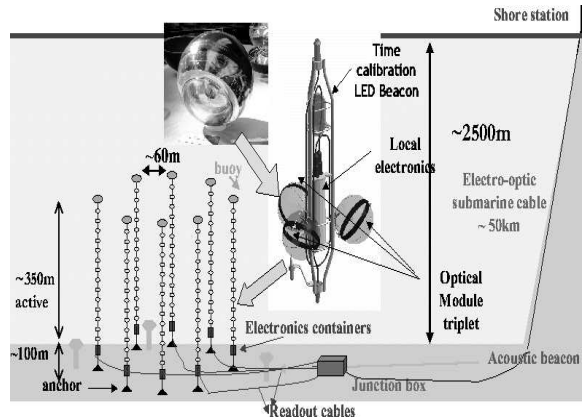}
\vskip -1 cm
\caption{Schematic view of the detector showing an OM and some
details of a storey enlarged.}
\label{fig1}
\end{figure}
\section{Detector design and physics studies}

The final detector design (Fig.~\ref{fig1}) 
was defined after the R\&D environmental 
and electronics studies and after simulations of the detector
response, given cost constraints.
Ten strings, holding 90 PMTs each,
made by mechanically strong electro-optical
cables and anchored at the sea bed will be deployed. 
Strings, kept taut by buoys at the top, will be separated by 
roughly 60 m.
Each string contains 30 storeys separated
by 12 m. Each storey is equipped with 3 OMs oriented outwards at $45^{\circ}$
downwards.
The 10'' Hamamatsu R7081-20 PMTs in the OMs are sensitive to single photons 
with transit time spread $<3$ ns (FWHM). 
The electronics chain has a timing precision of $\sim 1$ ns. 
Signals are digitized, then read out via
optical fibers in cables. A submarine will connect 
the cables from individual strings to a junction box at the end of 
an electro-optical cable sending signals to a 
shore base where data are recorded. The $\sim 40$ km long cable has been
recently successfully deployed. 
An extra-string will be devoted
to environmental parameter measurements.
The possibility to extend the detector up to 14 strings is considered.

The ANTARES scientific program is mainly devoted to the detection of
neutrinos of astrophysical origin with energies above 10 GeV.
Extra-galactic and galactic objects could be emitting neutrinos as
a consequence of hadronic production mechanisms alternative to 
the electromagnetic processes. 

\begin{figure}[htb]
\vspace{-.5cm}
\includegraphics[width=18pc, height=13pc]{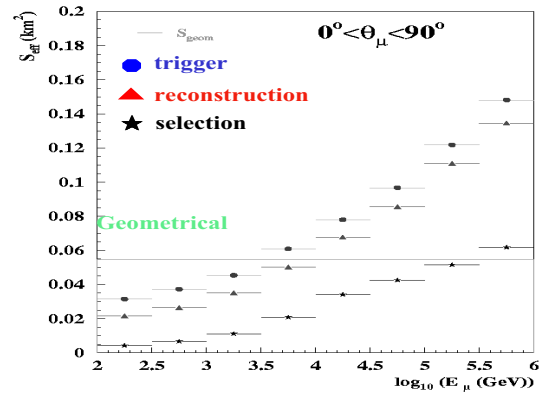}
\vspace{-1 cm}
\caption{Effective area as a function of simulated upward-going muon energy.
}
\label{fig2}
\end{figure}

Parameters to qualify a neutrino telescope are its effective area,
which includes reconstruction 
and selection efficiency~\footnote{At high energies the number of `volume' 
events with the vertex inside the
instrumented region is relatively small.}, 
the angular resolution and the energy resolution.
Simulation results show that the effective area exceeds the geometrical
surface in the high energy region of 
interest even after strict selections (see Fig.~\ref{fig2}).
The neutrino pointing resolution and the intrinsic angular resolution
of the telescope, defined as the median angular separation between the
'true' and the reconstructed muon track, has been studied by simulations.
For reconstructed events the angular resolution improves with energy.
Above 10 TeV, the pointing resolution is found to be within $\sim
0.2^{\circ}$, in spite of light scattering effects.

Below 100 GeV, neutrino energy is inferred from muon range
in the detector, while at energies $>1$ TeV an energy estimator
based on the PMT charge amplitudes currently allows a 
confidence interval ($68\%$ C.L.) on the measured energy $E$ between 
$\sim E/3$ and $\sim 3 E$. 
We are still working on rejection algorithms to suppress atmospheric 
neutrinos based on the deduced energy. Simulations show that by imposing a 
$\sim 10$ TeV cut, 20/150 events/yr are selected using single 
AGN \cite{Protheroe}/ 
diffuse AGN \cite{Stecker} fluxes with a background of $\sim 10$
atmospheric neutrino events/year.

Preliminary results on neutralino searches from the core of the Sun and
from the Galactic Center,
where these particles could be trapped if
they are the dark matter in the Galaxy, have been shown in
\cite{Bailey}. The ANTARES sensitivity may be better than 
current experiments, taking into account the different energy
thresholds. The algorithms developed for the oscillation studies,
also used for dark matter searches, show that 
ANTARES is expected to reach down to an energy threshold 
of 10 GeV (thanks to the inclusion of single-string reconstructed events)
with a median angular error of $2.6^{\circ}$ below 100 GeV.  

Studies on the ANTARES performance for neutrino oscillations
show that $\Delta m^{2}$ can be measured throughout
the region allowed by Super-Kamiokande with a precision of about $30\%$
in 3 years. (For maximum mixing, the region
$\Delta m^{2} \sim 3 \cdot 10^{-4}-6 \cdot 10^{-1}$ eV$^{2}$
could be excluded with $90\%$ C.L.).
The observables used are the zenith angle $\theta$ and $E/L$ ($L$
is the baseline which is almost proportional to $\cos\theta$) for
4000 upgoing muons/year. Backgrounds and systematic
errors are under study.

\end{document}